\documentclass[twocolumn,showpacs,preprintnumbers,superscriptaddress,amsmath,amssymb]{revtex4}
\usepackage{amsfonts}
\usepackage{amsmath}
\usepackage{amssymb}
\usepackage{graphicx}%
\usepackage{footmisc}
\begin{document}
\title{Scalar and tensor perturbations in vacuum inflation}
\author{Zhiqiang Huang}
\affiliation{State Key Laboratory of Magnetic Resonances and Atomic and Molecular Physics, Wuhan Institute of Physics and Mathematics,
Chinese Academy of Sciences, Wuhan 430071, China}
\affiliation{ University of the Chinese Academy of Sciences, Beijing 100049, China}
\author{Dongfeng Gao}
\affiliation{State Key Laboratory of Magnetic Resonances and Atomic and Molecular Physics, Wuhan Institute of Physics and Mathematics,
Chinese Academy of Sciences, Wuhan 430071, China}

\author{Qing-yu Cai}
\thanks{Corresponding author. Electronic address: qycai@wipm.ac.cn}
\affiliation{State Key Laboratory of Magnetic Resonances and Atomic and Molecular Physics, Wuhan Institute of Physics and Mathematics,
Chinese Academy of Sciences, Wuhan 430071, China}
\date{\today}

\begin{abstract}
Recently, it was proposed that a small true vacuum universe can inflate spontaneously, in principle.
Furthermore, there should be matter creation in vacuum inflation due to quantum fluctuations, and the matter created will
influence the inflation simultaneously.
In this paper,  scalar and tensor perturbations in this model are analyzed and confronted with recent observations. These  perturbations are derived and expressed with Hubble flow-functions.
By comparing our calculations with experimental results, we can determine all the parameters in
this model.
Finally, with the determined parameters, we compute the evolution of the matter density
and show that the matter produced in inflation roughly fits the observations at present.

\end{abstract}


\maketitle

\section{INTRODUCTION}\label{s1}
It has been widely believed that there was a fast expansion period of the early universe,
called inflation, which successfully solved several severe problems in the Big Bang
cosmology~\cite{1993ppc..book.....P}. Although the concept of inflation is successful in the standard
cosmology model, the origin and the power of the inflation are still unknown.
It was thought that the Higgs field might drive the inflation, but there are some problems in the Higgs
inflation model~\cite{2014JCAP06039F}, such as the naturalness problem~\cite{blt10}.
Recently, it was mathematically proved that the universe can be spontaneously created from
nothing, in principle~\cite{PhysRevD.89.083510}. For a small true vacuum universe, its quantum potential
can drive its inflation. When the universe becomes large enough, its accelerating expansion ends.
This vacuum inflation model does not need a matter field to drive inflation, which is different from
slow-roll inflation~\cite{Lidsey:1995np}. However, vacuum inflation should be developed because there
are some parameters to be determined. When the issue of matter creation during inflation is considered, the vacuum
inflation derived by the quantum potential should be modified slightly. Furthermore, the quantum
fluctuations of the early universe, viewed as seeds for the growth of the structure of the universe, 
should also be given~\cite{ellis1999cosmological}.

This paper is organized as follows. Sec. \ref{s2}, we first briefly review the vacuum inflation deriven by the quantum potential and then investigate the modification of the inflation by the matter created in the inflation. 
In Sec. \ref{s3}, we calculate the scalar perturbations up to second-order perturbations of the Hamiltonian for the
scalar field and gravity in ADM form, without considering the concrete form of the scalar fields.
Then, we discuss the quantization of gravity waves and calculate the tensor perturbations.
In Sec. \ref{ravtm}, by comparing our calculations with experimental observations, we determine all the necessary parameters in the model.
In Sec. \ref{PC}, the matter density is calculated. With the previously determined parameters, we fit the value of matter density which is compatible with current observations.
Finally, we give discussions and conclusions in Sec. VI.

For simplicity, we use Planck units in this paper.

\section{Vacuum inflation and matter correction}\label{s2}

We first briefly review the vacuum inflation derived by the quantum potential and then
show the modification of the vacuum inflation by the matter created in inflation.

\subsection{Vacuum inflation}

There is a long-standing debate over whether a vacuum can inflate spontaneously~\cite{Vilenkin1984Quantum,Universe1984The,PhysRevD.58.023501}.
If the universe indeed came from a vacuum, the problem of creation in the Big Bang model can
be avoided~\cite{edp73}.
Recently, it was proved by the minisuperspace model that a small true vacuum bubble can
inflate because of its quantum potential~\cite{PhysRevD.89.083510}, which makes it possible
to construct a theory in which the universe grows up spontaneously.
For a true vacuum bubble, its action $S_{\text{gr}}$ can be written as
\begin{equation} \label{grp}
S_{\text{gr}}=\int \frac{1}{16{\pi}}\mathbb{R}\sqrt{-g}d^4x,
\end{equation}
where $\sqrt{-g}$ is the square root of the absolute value of the determinant of the metric, $\mathbb{R}$ is the curvature and $d^4x=d\tau d^3x$. $\tau$ is the conformal time defined as $\tau \equiv \int dt/a$, where $a$ is the scale factor.
Because the vacuum bubble is homogeneous and isotropic, the metric can be expressed by one parameter,
the scale factor $a$,
\begin{align} \label{ccs}
g_{\mu \nu}=\left(\begin{array}{cc}
a^2 & 0 \\
0 & -\frac{a^2}{1-K r^2}\delta_{ij} \\
\end{array}\right).
\end{align}
By substituting the metric above into Eq.~(\ref{grp}), it is easy to obtain the Hamiltonian of
this model $\mathfrak{H}=-l^{-2}\left(\frac{1}{4}l^4P_a^2+K a^2\right)$,
where $P_a=-2l^{-2}a'$ and $l^2\equiv 8 \pi/3$. Here, the primer  denotes the derivative
with respect to the conformal time. The parameter $K=1, 0, -1$ represents closed, flat,
and open bubbles, respectively.
We adopt $K=0$  as default for all calculations and only keep the $K$ explicitly in some formula for the purpose of comparing  with classical case.
In the minisuperspace model, the momentum operator is defined as
\begin{equation} \label{can}
P_a^2=-a^{-p}\partial_a(a^p\partial_a),
\end{equation}
where $p$ is the factor of ordering and denotes the ambiguity of rules for quantization.
Because there is no complete quantum theory for gravity, we treat $p$ as an uncertain
parameter.
In principle, the cosmological wavefunction satisfies the Wheeler-DeWitt equation, $\mathfrak{H}\psi(a)=0$.
Decomposing the cosmological wavefunction into two real functions $\psi(a)=R(a)\exp(i S(a))$ and using Eq.~(\ref{can}), we have 
$\frac{1}{4}l^4(\partial_a S)^2+K a^2+Q=0$, and
\begin{equation} \label{qpt}
4l^{-4}Q=-\frac{\partial_{aa} R}{R}-\frac{p}{a} \frac{\partial_a R}{R},
\end{equation}
which is called the quantum potential.
Using the guidance relation ~\cite{PhysRevD.89.083510}
\begin{equation} \label{gd}
\partial_{a'}\mathcal{L}=-2l^{-2}a'=\partial_a S,
\end{equation}
one can obtain 
\begin{equation}
H^2=-\frac{Q}{a^4}-\frac{K}{a^2},\notag
\end{equation}
where $H$ is the Hubble parameter. It is often defined as $d\ln a/d t$.
The $\mathcal{L}$ in the guidance relation is the the Lagrangian of Eq.~(\ref{grp}). The formula of the Hubble parameter above shows that the quantum potential $Q$
has the capacity to drive the vacuum bubble towards accelerating expansion.

\begin{figure}[ht]
\centering
\includegraphics[width=0.45\textwidth]{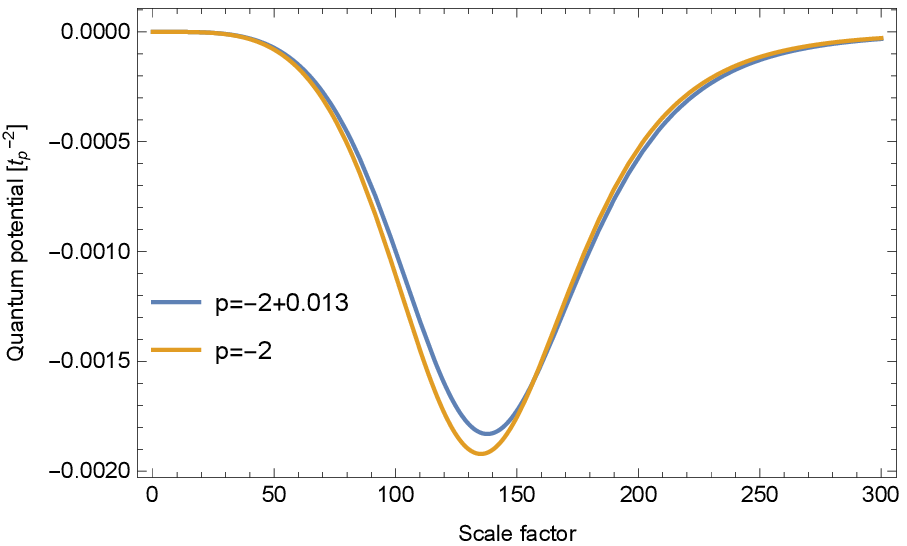}
\includegraphics[width=0.45\textwidth]{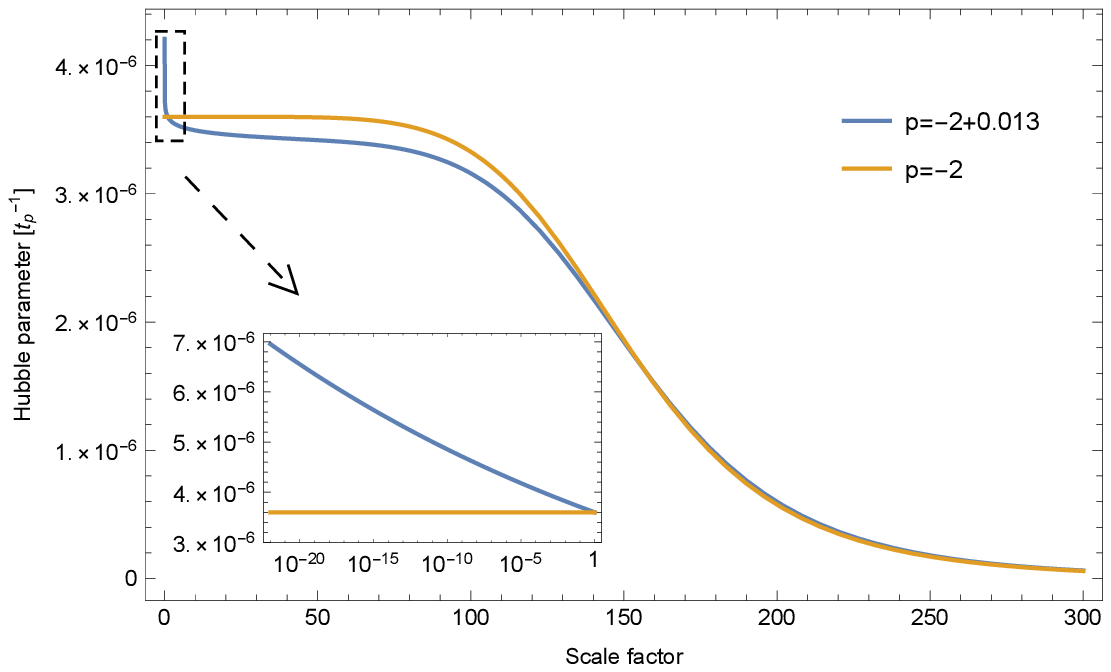}
\caption{The quantum potential $Q$ and Hubble parameter $H$ vs the scale factor $a$ with different $p$. They are from Eq.~(\ref{qp}) and  Eq.~(\ref{h}) without matter correction.
We use the parameter $C_1/C_2$ fixed in Sec. \ref{ravtm} to get these exact results. When comparing it with asymptotic expression Eq.~(\ref{H}), we can easily find when Eq.~(\ref{H}) is approximately right. The late time of this plot means that the universe starts to exit the inflation. The quantum effect does decrease strongly, but is still much bigger than the influence of matter in that time.}
\label{cmqh}
\end{figure}

The analytical formula of the wavefunction can be obtained by solving the
Wheeler-DeWitt equation, $\psi(a)=iC_1a^{1-p}/(1-p)-C_2$, where $p\neq1$ and
$C_1$ and $C_2$ are arbitrary complex numbers. In order to get an inflation solution, we set $C_1$ and $C_2$ as real numbers. Because only the value of $|1-p|$ is crucial for the inflation solution, we set $p<1$  in the paper for convenience. 
Then, we have $S=\arctan[-C_1a^{1-p}/C_2(1-p)]$~\cite{PhysRevD.89.083510}.
With the guidance relation Eq.~(\ref{gd}), when $|\frac{C_1}{C_2(1-p)}a^{1-p}|^2\ll1$, we can obtain
the asymptotic form of the Hubble parameter
\begin{equation}\label{H}
H=\frac{l^2}{2}\frac{C_1}{C_2}a^{-2-p}.
\end{equation}
We can see how small $a$ should be when we determine $C_1$, $C_2$ and $p$. From the results of Sec. \ref{ravtm}, this equation is roughly right when $a<50$. We can also get 
this conclusion from Fig.~(\ref{cmqh}), where $H$ is almost constant till $a\sim50$. 

Now, it is easy to obtain the time-dependent evolution for the scale factor,
\begin{equation} \label{at}
a(t)=\begin{cases}
\left[\frac{l^2}{2}\frac{C_1}{C_2}(2+p)(t-t_0)\right]^{\frac{1}{2+p}}             &{p\neq-2}\\
e^{\frac{l^2}{2}\frac{C_1}{C_2}(t-t_0)}          &{p=-2}
\end{cases}.
\end{equation}
When $p=-2$, we have an exponential inflation solution.
When $-2<p<-1$, we get power-law inflation.

In Fig.~(\ref{cmqh}), we show the evolutions of the quantum potential and
the Hubble parameter with the variable of the scale factor $a$ for a specific ordering
factor $p=-2$.
The Hubble parameter is nearly constant in the early stage of the universe, which is quite similar to slow-roll inflation. After a while, the Hubble parameter decreases rapidly to zero, which means that the inflation can exit spontaneously.
The quantum potential is the power of the vacuum inflation and plays the role of the scalar field in the
slow-roll inflation model~\cite{PhysRevD.51.2729}.
We also find that different values of $p$ have different quantum potentials, which leads
to different evolution of the universe.

\subsection{Matter correction}\label{mc}

There are quantum fluctuations such as the creation and annihilation of virtual particles
even in the vacuum of space. Due to the accelerating expansion, the virtual particles may be
separated before annihilation and then become real particles. Therefore, there is matter creation
during vacuum inflation~\cite{Modak:2012cj}. With the matter creation, the action of the early universe is
modified by a scalar field,
\begin{equation} \label{tts}
S_{\text{gr}}+S_{\text{SC}}=\int \left[\frac{1}{16{\pi}}\mathbb{R}+
\left(\frac{1}{2}\varphi _{,\alpha }\varphi ^{,\alpha }
-V(\varphi )\right)\right]\sqrt{-g}d^4x,
\end{equation}
where $V(\varphi)$ is the potential of the scalar field.
Here, we don't need to restrict the form of the potential of the scalar field,
which is different to the slow-roll inflation theory.
By substituting Eq.~(\ref{ccs}) into Eq.~(\ref{tts}),
we can obtain the Hamiltonian of the system
\begin{equation} \label{wave}
\mathfrak{H}=-l^{-2}\left(\frac{1}{4}l^4P_a^2+K a^2\right)+\left(\frac{1}{2}
\frac{P_{\varphi }^2}{a^2}+\frac{1}{2} a^4 \varphi _{,i}^2+a^4 V(\varphi) \right),
\end{equation}
where $P_{\varphi }=a^2 \varphi'$ and $P_{a}$ was defined in Eq.~(\ref{can}).
the wave function of the universe still satisfies the Wheeler-DeWitt equation,
$\mathfrak{H}\psi(a,\varphi)=0 $, but it is difficult to obtain analytical solutions for
this equation.
For the vacuum inflation model, the inflation is driven by the quantum effects
of the universe, particularly at the beginning of the early universe. Under this condition,
the wavefunction of the universe can be approximately written as
$\psi(a,\varphi)=\psi_a(a)\psi_s(a,\varphi)$, where $\psi_a(a)$ is the wavefunction for the
space of the universe and $\psi_s(a,\varphi)$ is the wavefunction of the scalar field.
The quantity of $P_a\psi_s(a,\varphi )$ is negligible compared with $P_a\psi_a(a)$.
In this way, Eq.~(\ref{wave}) can be simplified as
\begin{align}
\label{mands}
&\left(\frac{1}{2}\frac{P_{\varphi }^2}{a^2}+\frac{1}{2} a^4 \varphi _{,i}^2+a^4 V(\varphi)
\right)\psi_s(a,\varphi )=a^4 \rho(a)\psi_s(a,\varphi ), \notag \\
&\left(-l^{-2}\left(\frac{1}{4}l^4P_a^2+K a^2\right)+a^4 \rho(a)\right)\psi_a(a)=0,
\end{align}
where $\rho(a)$ can be treated as matter density, which we will explain using Eq.~(\ref{h}).
By expressing $\psi_a(a)$ as two real functions $\psi_a(a)=R_a(a)\exp(i S_a(a))$ and
substituting Eq.~(\ref{can}) into Eq.~(\ref{mands}), we can obtain
\begin{equation}
\frac{1}{4}l^4(\partial_a S)^2+K a^2+Q=l^2 a^4 \rho(a),\notag
\end{equation}
where $Q$ is determined by Eq.~(\ref{qpt}).
Using guidance relation Eq.~(\ref{gd}) we find that the Hubble parameter takes the form
\begin{equation} \label{h}
H^2=-\frac{Q}{a^4}-\frac{K}{a^2}+\frac{8 \pi   }{3}\rho.
\end{equation}
Comparing Eq.~(\ref{h}) with the classical Friedmann equation~\cite{fridmann} when $Q\rightarrow0$, we know that $\rho(a)$ is exactly the matter density.

From the first equation of Eqs.~(\ref{mands}), we have $\mathfrak{H}_\varphi=a^4\rho(a)$, where $\mathfrak{H}_\varphi$ is the $\varphi$ part of the $\mathfrak{H}$. The derivative of the Eq.~(\ref{h}) with respect to the normal time can give
\begin{equation} \label{eae}
	2\dot{H}=3\frac{Q}{a^4}-\frac{1}{a^3}\frac{d Q}{ d a}+K-\frac{8 \pi   }{3}\frac{\mathfrak{L_\varphi}}{a^4},
\end{equation}
where $\mathfrak{L_\varphi}\equiv P_\varphi{}^2 -\mathfrak{H}_\varphi$ . Eq.(\ref{eae}) is the Einstein acceleration equation of this model. When $Q\rightarrow0$, it goes back to the classical Einstein acceleration equation.

Because vacuum inflation is essentially driven by the quantum potential, the contribution
of matter to the evolution of the early universe can be neglected.
With Eqs.~(\ref{qpt}) and (\ref{at}), the quantum potential of the early universe can be
written as
\begin{equation} \label{qp}
Q=-\frac{l^4}{4}\frac{C_1^2 C_2^2 (p-1)^4 a^{2 p}}{\left(C_2^2 (p-1)^2 a^{2 p}+a^2 C_1^2\right)^2},
\end{equation}
where $C_1$ and $C_2$ are two parameters to be determined.

\begin{figure}[ht]
\centering
\includegraphics[width=0.45\textwidth]{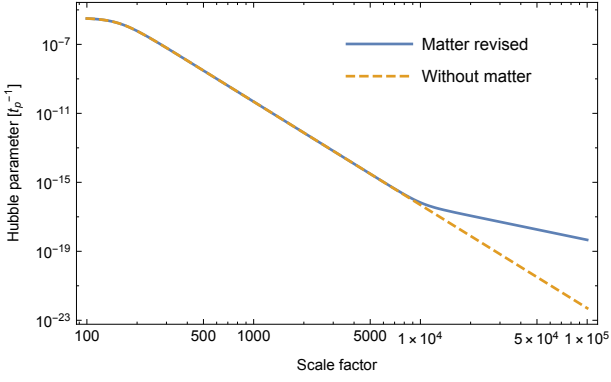}
\caption{Comparison of the Hubble parameter with and without the matter
revision in Eq.~(\ref{h}) in the late stage of inflation and after. The figure uses the parameter fixed in Sec. \ref{ravtm}. We can see that the contribution of matter becomes more and more important. The range 
$a>10^4$ indicates the end of the inflation.}\label{cm}
\end{figure}

As discussed above, there are energy fluctuations as the virtual particles are created and annihilated.
The virtual particles may become real particles probabilistically when tunneling through the cosmological
horizon~\cite{Kumar2012Hawking}.
If we consider Eq.~(\ref{mands}) in de Sitter space, the radius $r$ of the horizon is $1/H$,
and the temperature of radiation from the horizon is $H/(2\pi)$~\cite{Modak:2012cj}.
Using the Stefan-Boltzmann law, the creation rate for the matter density is
\begin{equation}
\dot{\rho }=\frac{4 j \text{$\pi $r}^2}{\frac{4 \text{$\pi $r}^3}{3}}=\frac{3 \text{$\sigma $H}^5}{16 \pi ^4},
\end{equation}
where the dot denotes derivative with respect to the normal time $t$, $j$ is the energy flux from the horizon,
and $\sigma$ is the Stefan-Boltzmann constant.
The particles created during inflation are relativistic due to the high temperature at this stage, so
the mass-conservation equation gives $\dot{\rho }=-4 H \rho$.
Considering both creation and dilution of matter simultaneously, we have 
\begin{equation} \label{rho}
\dot{\rho }=\frac{3 \text{$\sigma $H}^5}{16 \pi ^4}-4 H \rho.
\end{equation}
Combining Eqs.~(\ref{h}) and (\ref{rho}), we can obtain the equation for the change of the matter density
with the growth of the early universe.
The evolution of the Hubble parameter can also be obtained, which is numerically shown in Fig.~(\ref{cm}).
When the universe is small, the Hubble parameter is modified slightly by the matter created during inflation, which is the previous assumption that the inflation is mainly driven by its quantum potential.
When the early universe becomes large, the contribution of matter to inflation is relatively large,
whereas the dynamics of the universe are classical because $Q\rightarrow0$ for a large enough universe~\cite{fridmann}. 

\section{Cosmological perturbations}\label{s3}


The measurement of CMBR shows that there were quantum fluctuations of gravity during inflation,
which were the seeds of the galaxies of our universe.
With the results of CMBR, we can look into the details of inflation and verify
the validity of the inflation model.
Although all kinds of fields can generate corresponding cosmology perturbations, we only consider scalar and tensor fluctuations during inflation, because they can be determined by the measurement of CMBR.
In the following, we will first follow the standard cosmological perturbation theory but express the scalar and tensor perturbations with the Hubble parameter and its derivative. Then, we will show that it is right to use these classical conclusions in this model.

From the calculations, we see that the Hamiltonian of perturbations can be expressed by the Hubble parameter and its derivative.
The perturbations are decided by how the universe inflated rather than the concrete form of the scalar
fields. In this model, quantum potential, together with field perturbations leads to cosmological perturbations. It is different from the standard model of inflation, where the scalar field and its perturbations lead to cosmological perturbations.

\subsection{Scalar perturbations}\label{sp}

Scalar perturbations are gravity perturbations caused by fluctuations of scalar fields.
We need the four quantities $B$, $\psi$, $E$ and $\phi$ to describe the most general form of
scalar metric perturbations
\begin{align}
g_{ij}=\left(\begin{array}{cc}
\mathcal{N}^2-\mathcal{N}_i\mathcal{N}^i & -\mathcal{N}_i \\
-\mathcal{N}_i & -\gamma _{ij} \\
\end{array}\right),
\end{align}
where $\mathcal{N}_i=a^2\partial_i B $, $ \gamma _{ij}=a^2(1-2\psi )
\delta _{ij}+2a^2\partial_{i}\partial_{j}E $
and $\mathcal{N}=a\left(1+\phi -\frac{1}{2}\phi^2+\frac{1}{2}\partial_{i}B\partial_{i}B\right)$.
Furthermore, we use the quantity $\varphi_0 +\delta \varphi$ to describe perturbations of the scalar field.

\begin{figure}[ht]
\centering
\includegraphics[width=0.45\textwidth]{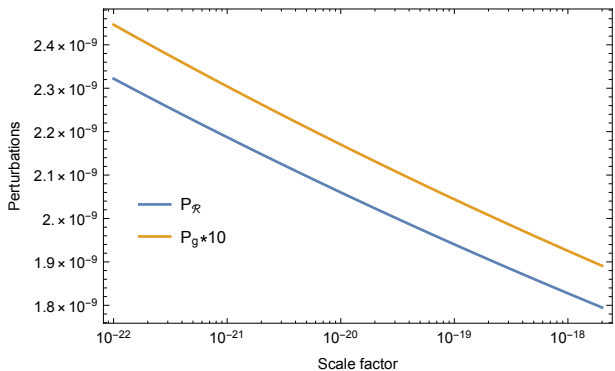}
\caption{The power spectrum of the curvature perturbation and gravitational waves in this model. The figure uses the parameter fixed in Sec. \ref{ravtm}.} \label{prpic}
\end{figure}

The scalar perturbations can be tested by the intrinsic curvature perturbation of
the comoving hypersurface $\mathcal{R}$, which is a gauge invariance quantity,
\begin{equation}
\mathcal{R}\equiv-\psi -\frac{ \mathcal{H} }{\partial_\tau \varphi}\delta \varphi,
\end{equation}
where $\mathcal{H}\equiv a'/a$. 
Using the back ground equations to cancel the background action,
the first-order perturbations and the background scalar field $\varphi_0$, we can simplify the action
in Eq.~(\ref{tts}) to the second-order ~\cite{Mukhanov1992203}
\begin{align}
\begin{split}
&\delta _2S=\frac{1}{6\ell ^2}\int \left\{a^2\left[-6(\psi ')^2-12\mathcal{H}\phi \psi '-2\left(\mathcal{H}'
+2\mathcal{H}^2\right)\phi^2 \right. \right. \\
&-2\psi _{,i}\left(2\phi _{,i}-\psi _{,i}\right)+3\ell ^2\left((\delta \varphi ')^2-\delta \varphi _{,i}\delta
\varphi _{,i}-V_{,\varphi \varphi}a^2\delta \varphi ^2\right)\\
&+3\ell ^2\left(\varphi _0'(\phi +3\psi )'\delta \varphi '-2V_{,\varphi }a^2\phi \delta \varphi \right)\left. \right]\\
&\left.+2(B-E')_{,ii}\left(3\ell^2\varphi _0'\delta \varphi -2\psi '-2\mathcal{H}\phi \right)\right\}d^4x.
\end{split}
\end{align}
Furthermore, we choose a simple gauge as $\delta \varphi=0$ and $E=0$ to eliminate $B$ and $\phi$.
A straightforward calculation gives the action that only contains one field,
\begin{equation}\label{final perturbations}
\delta _2S=\frac{1}{6\ell ^2}\int -2a^2\frac{\dot{H}}{H^2}\left[(\psi ')^2-\psi _{,i}{}^2\right]d^4x,
\end{equation}
where we use both the normal time and the conformal time to simplify our calculations. $\dot{H}$ is $d H/d t$ and $\psi '$ is  $d\psi/d\tau$.
For convenience, we introduce two variables
$z=\sqrt{-a^2\dot{H}/H^2}=a\sqrt{-\dot{H}}/H$
and $u=-z\mathcal{R}/(2\sqrt{\pi})$ in our calculations.
Varying Eq.~(\ref{final perturbations}) and using Fourier transformation in the space variables, we can obtain
\begin{equation} \label{eq}
\left(\frac{d^2}{d\tau ^2}+k^2-\frac{1}{z}\frac{d^2z}{d\tau ^2}\right)u_k=0,
\end{equation}
where $u_k$ is the image function of $u$.
For easy comparison with experimental results, we use the Hubble flow-functions (HFFs)~\cite{Schwarz2001243} to simplify our calculations,
\begin{align} \label{hff}
&\epsilon _1=-\frac{ d\ln H }{ d\ln a },
&\epsilon _{i+1}=\frac{ d\ln \epsilon _i}{d\ln a},
\end{align}
where $i\geq 1$. Expressing $z$ with HFFs, we obtain $z=a\sqrt{\epsilon _1}$ and
\begin{equation} \label{zpp}
\frac{1}{z}\frac{d^2z}{d\tau ^2}=a^2 H^2 \left(\frac{\epsilon _2^2}{4}+\frac{3 \epsilon _2}{2}-\frac{\epsilon _1
\epsilon _2}{2}+\frac{\epsilon _3 \epsilon _2}{2}-\epsilon _1+2\right).
\end{equation}
Integrating the conformal time $\tau =\int 1/a \, dt=\int 1/(a^2 H) \, da$ by parts repeatedly gives 
\begin{align} \label{tau}
&\tau =-\frac{\sum _{i=0}^{\infty } f_i}{\text{aH}},
&f_{i+1}=f_i \left(\frac{d\ln f_i }{d\ln a}+\epsilon _1\right),
\end{align}
where $f_0=1$.
Combining Eqs.~(\ref{zpp}) and (\ref{tau}), we obtain
\begin{align} \label{zpps}
\begin{split}
\frac{1}{z}\frac{d^2z}{d\tau ^2}&=\frac{\left(\sum _{i=0}^{\infty } f_i \right) ^2\left(\frac{\epsilon _2^2}{4}+\frac{3
\epsilon _2}{2}-\frac{\epsilon _1 \epsilon _2}{2}+\frac{\epsilon _3 \epsilon _2}{2}-\epsilon _1+2\right)}{\tau^2} \\
&=\frac{\left(\left( \frac{3}{2} +F(\epsilon_1,\epsilon_2,...) \right)^2-\frac{1}{4}\right)}{\tau ^2},
\end{split}
\end{align}
where $F(\epsilon_1,\epsilon_2,...)$ represents the multinomial of $\epsilon_1$, $\epsilon_2$, \dots, and does not
contain a constant term. In this model, HFFS and $F(\epsilon_1,\epsilon_2,...)$ is constant during the early stage of inflation. We will show that in  Sec. \ref{ravtm}.
With Eq.~(\ref{zpps}) and constant $F(\epsilon_1,\epsilon_2,...)$, we can solve Eq.~(\ref{eq}) analytically.
In the limit $k/(aH)\to\infty$, $u_k$ has a normalized solution as $ u_k(\tau )\to -e^{-\text{i}k\tau }/\sqrt{2 k}$.
Thus, the solution of Eq.~(\ref{eq}) is
\begin{equation}
u_k(\tau )=\frac{\sqrt{\pi}}{2} e^{\frac{1}{2} \pi  i \left(\nu +\frac{1}{2}\right)}\sqrt{-\tau }  H_\nu(-k\tau), \notag
\end{equation}
where $H_\nu$ is the Hankel function of the first kind with $\nu\equiv \frac{3}{2} +F(\epsilon_1,\epsilon_2,...)$. We can then calculate the vacuum fluctuations of
the curvature perturbation $\mathcal{R}$. The power spectrum $\mathcal{P}_{\mathcal{R}}(k)$ is defined
in terms of the expectation value of inflation state $\left\langle \mathcal{R}_\mathbf{k} \mathcal{R}_\mathbf{l}{}^*\right\rangle \equiv 2 \pi ^2 \mathcal{P}_{\mathcal{R}}
\delta ^3 (\mathbf{k}-\mathbf{l})/k^3 $,
where $\mathcal{R}_\mathbf{k}$ is defined as
\begin{equation}
	\mathcal{R}=\int \frac{d^3\mathbf{k}}{(2 \pi )^{3/2}}\mathcal{R}_\mathbf{k}(\tau) e^{i \mathbf{k}\cdot \mathbf{x}}.\notag
\end{equation}
Considering the quantization of $u_k$, we have $\left\langle \mathcal{R}_
\mathbf{k} \mathcal{R}_\mathbf{l}{}^*\right\rangle = 4\pi\left\langle u_\mathbf{k} u_\mathbf{l}{}^*\right\rangle/z^2=4\pi
|u_k|^2\delta ^3 (\mathbf{k}-\mathbf{l})/z^2$.
Finally, we can obtain
\begin{equation} \label{pr}
\mathcal{P}_{\mathcal{R}}^{1/2}(k)=\frac{1}{\sqrt{\pi}}2^{\nu-3/2}\frac{\Gamma(\nu)}{\Gamma(3/2)}(\sum _{i=0}^{\infty }
f_i)^{-\nu+1/2}\frac{H}{\sqrt{\left|\epsilon _1\right|}} \Bigg\lvert _{k=aH}.
\end{equation}
The scalar perturbations do not depend on special scalar fields.
All scalar fields in the fundamental theory contribute a part of the scalar perturbations $\mathcal{P}_{\mathcal{R}}(k)$.
In the standard model, we have scalar fields $\Phi$ with complex doublets.
Therefore, there are two scalar field contributions to the scalar perturbations~\cite{Marco2005Slow}, and the total power spectrum
$\mathcal{P}_{\mathcal{R}}^\prime(k)=2\mathcal{P}_{\mathcal{R}}(k)$. We show how curvature perturbations change
along with the scale factor in Fig.~(\ref{prpic}). The perturbations are close to constant in the early stage of inflation.

\subsection{Gravitational waves}\label{gw}
Gravitational waves are gravitational perturbations that are produced by the vacuum fluctuations of gravity.
The linear tensor perturbations can be written as $g_{\mu \nu }=a^2(\tau)
\left[\eta _{\mu \nu }+h_{\mu \nu }\right]$.
In the transverse traceless gauge as $h_0=h_{0 i}=\partial ^ih_{ij}=h_{ii}=0$,
there are two independent states~\cite{Weinberg1972Gravitation}.
The action of the perturbations can be expressed as
\begin{equation} \label{gravity}
S_g=\frac{1}{64 \pi }\int a^2(\tau) \partial _{\mu } h^i{}_j \partial ^{\mu }h_i^jd^4x.
\end{equation}
It is convenient to calculate the perturbations with rescaled variables $v_\lambda(x)=(32\pi)^{-1/2}a(\tau)h_\lambda(x)$,
where $h_\lambda(x)$ and $\lambda=+,\times$ represent two independent states.
By varying Eq.~(\ref{gravity}) and using Fourier transformation in the space variables, we can obtain
\begin{equation}
\frac{d^2\nu _k}{d\tau ^2}+\left(k^2-\frac{1}{a}\frac{d^2a}{d\tau ^2}\right)\nu _k=0,
\end{equation}
where $v_k$ is the image function of $v$.
Using of Eq.~(\ref{hff}), we can simplify the following expression
\begin{equation}
\frac{1}{a}\frac{d^2a}{d\tau ^2}=2a^2H^2\left(1-\frac{1}{2}\epsilon_1\right).
\end{equation}
Substituting Eq.~(\ref{tau}) into the equation above, we can obtain
\begin{align} \label{apps}
\begin{split}
\frac{1}{a}\frac{d^2a}{d\tau ^2}&=\frac{\left(\sum _{i=0}^{\infty } f_i \right) ^2\left(1-\frac{1}{2}\epsilon_1\right)}{\tau^2} \\
&=\frac{\left(\left( \frac{3}{2} +G(\epsilon_1,\epsilon_2,...) \right)^2-\frac{1}{4}\right)}{\tau ^2},
\end{split}
\end{align}
where $G(\epsilon_1,\epsilon_2,...)$ represents the multinomial of $\epsilon_1$, $\epsilon_2$, \dots 
and does not contain a constant term.
Repeating the same procedure as the case of scalar perturbations, we find that
$v_k(\tau )$ has the same solutions as that of $u_k(\tau )$ except
that $\nu$ is replaced by $\mu$, and $\mu$ is defined as $\mu= \frac{3}{2} +G(\epsilon_1,\epsilon_2,...)$.

The spectrum of gravitational waves $\mathcal{P}_{g}(k)$ is defined as
$\left\langle h_{\mathbf{k},\lambda} h_{\mathbf{l},\lambda} {}^*\right\rangle
\equiv 2 \pi ^2 \mathcal{P}_{g}\delta ^3 (\mathbf{k}-\mathbf{l})/k^3 $~\cite{Lidsey:1995np}.
Using the relation between $v_\lambda(x)$ and $h_\lambda(x)$, we can obtain
\begin{equation}
\left\langle v_{\mathbf{k},\lambda} v_{\mathbf{l},\lambda} {}^*\right\rangle
\equiv \frac{1}{32\pi} a(\tau)^2 2 \pi ^2 \mathcal{P}_{g}\delta ^3 (\mathbf{k}-\mathbf{l})/k^3.
\end{equation}
With the quantization of $v_{\mathbf{k},\lambda}$, we have that $\left\langle v_{\mathbf{k},\lambda} v_{\mathbf{l},
\lambda'}{}^*\right\rangle=|v_{\mathbf{k},\lambda}|^2\delta ^3 (\mathbf{k}-\mathbf{l})
\delta _{\lambda ,\lambda '}$.
Furthermore, we can obtain the spectrum of the gravitational waves as
\begin{equation} \label{pg}
\mathcal{P}_{g}^{1/2}(k)=\frac{2}{\sqrt{\pi }}2^{\mu-1/2}\frac{\Gamma(\mu)}{\Gamma(3/2)}
(\sum _{i=0}^{\infty } f_i)^{-\mu+1/2}H\Bigg\lvert _{k=aH}.
\end{equation}
In Fig.~(\ref{prpic}), we show the change of the power spectrum of gravitational waves with the scale factor.
The power spectrum is also close to a constant in the early stage of inflation.
It decreases in the same way as the Hubble parameter does when the scale factor becomes large.
The perturbations is nearly constant when they leave the horizon, all the calculation and experimental observations are based on $k=aH$~\cite{Weinberg1972Gravitation}.

\subsection{Perturbations and quantum potential}
The calculation of Sec. \ref{sp} and Sec. \ref{gw} are totally classical.
This model is based on quantum theory.
It seems unreasonable to use Eq.~(\ref{final perturbations}) in this model. 
We need to discuss how to get a reasonable perturbation theory in this model. 
The derivation satisfies the following principle. 
First, all the variables of the Hamiltonian are operator. They may not communicate with each other.
Second, the operator must satisfy the classical equations of motion.
Third, when the operator acts on the wavefunction which is not the eigenstate of this operator, we need add the corresponding quantum potential in and replace the operator with its classical definition. 
For example, we take
\begin{align} \label{of}
	\begin{split}
		\hat{H}^2\psi_a(a)=\left(H^2+\frac{Q}{a^4}\right)\psi_a(a),\\
		\hat{\dot{H}}\psi_a(a)=\left(\dot{H}+\frac{3}{2}\frac{Q}{a^4}-\frac{1}{2a^3}\frac{d Q}{d a}\right)\psi_a(a).\\
	\end{split}
\end{align}
With Eq.~(\ref{of}), we can easily obtain Eq.~(\ref{h}) from Eq.~(\ref{mands}).

First, we discuss the gravitational waves. The
wave function of the universe satisfies 
\begin{equation} \label{gwn}
	\left(\mathfrak{H}(\hat{a},\hat{H},\hat{\varphi})+\mathfrak{H}_g(\hat{a},\hat{h})\right)\psi(a,\varphi,h)=0,
\end{equation}
where $\mathfrak{H}(\hat{a},\hat{\varphi})$ is defined by Eq.~(\ref{wave}) and $\mathfrak{H}_g(\hat{a},\hat{h})$ is the Hamiltonian of  Eq.~(\ref{gravity}). 
We can separate the wave function similar to that of the Sec. \ref{mc} and obtain $\psi(a,\varphi,h)=\psi_b(a,\varphi)\psi_h(a,\varphi,h)$. With that we can simplify Eq.~(\ref{gwn}) to
\begin{align}
	\begin{split}
		\mathfrak{H}_g(a,\hat{h})\psi_h(a,\varphi,h)=a^4 \rho_h(a)\psi_h(a,\varphi,h),\\
		\left(\mathfrak{H}(\hat{a},\hat{\varphi})+a^4 \rho_h(a)\right)\psi_b(a,\varphi)=0.
	\end{split}
\end{align}
Comparing it with Eq.~(\ref{mands}), we can know that $\rho_h(a)$ is the energy of gravitational waves and $\psi_b(a,\varphi)$ is the background wave function of an inflation universe which contains the correction of gravitational waves.  $\psi_b(a,\varphi)$ represents gravitational waves at the scale factor $a$.
As the energy of perturbations is negligible, the $\psi_b(a,\varphi)$  approximates to the  $\psi(a,\varphi)$ of the Sec. \ref{mc}. The $\psi_h(a,\varphi,h)$ goes back to the calculation of Sec. \ref{gw}.

Second, we discuss the scalar perturbations. The Hamiltonian is 
\begin{equation}
	\mathfrak{H}(\hat{a},\hat{H},\hat{\varphi})+\mathfrak{H}_s(\hat{a},\hat{H},\hat{\dot{H}},\hat{\phi} ,\hat{\psi} ,\hat{B},\hat{E},\hat{\delta\varphi}),
\end{equation}
where $\mathfrak{H}_s(\hat{a},\hat{H},\hat{\phi} ,\hat{\psi} ,\hat{B},\hat{E},\hat{\delta\varphi})$ is the total Hamiltonian of scalar perturbations. Under the second principle, We can use the Hamiltonian of Eq.~(\ref{final perturbations}) which means the wave function satisfies
\begin{equation}\label{spn}
	\left(\mathfrak{H}(\hat{a},\hat{H},\hat{\varphi})+\mathfrak{H}_\mathcal{R}(\hat{a},\hat{H},\hat{\dot{H}},\hat{\psi})\right)\psi(a,\varphi,\psi)=0,
\end{equation}
where $\mathfrak{H}_\mathcal{R}$ is the Hamiltonian of Eq.~(\ref{final perturbations}). We can separate the wave function similar to that of the Sec. \ref{mc} and obtain $\psi(a,\varphi,\psi)=\psi_b(a,\varphi)\psi_\mathcal{R}(a,\varphi,\psi)$. With that we can simplify Eq.~(\ref{spn}) to
\begin{align}
	\begin{split}
		\mathfrak{H}_\mathcal{R}(a,H,\dot{H},\hat{\psi})\psi_\mathcal{R}(a,\varphi,\psi)=a^4 \rho_\mathcal{R}(a)\psi_\mathcal{R}(a,\varphi,\psi),\\
		\left(\mathfrak{H}(\hat{a},\hat{H},\hat{\varphi})+a^4 \rho_\mathcal{R}(a)\right)\psi_b(a,\varphi)=0.
	\end{split}
\end{align}
compared with Eq.~(\ref{mands}),
the $\rho_\mathcal{R}(a)$ is the energy of scalar perturbations and $\psi_b(a,\varphi)$ is the background wave function of an inflation universe which
contains the correction of scalar perturbations. $\psi_\mathcal{R}(a,\varphi,\psi)$ represents scalar perturbations at the scale factor $a$. Since the Hubble parameter has been determined for the given $a$ in this model, the $H$ and $\dot{H}$ in $\mathfrak{H}_\mathcal{R}$ will not bring any vagueness. As
the energy of perturbations is negligible, the $\psi_b(a,\varphi)$ here approximates to the  $\psi(a,\varphi)$ of the Sec. \ref{mc}. The $\psi_\mathcal{R}(a,\varphi,\psi)$ goes back to the calculation of Sec. \ref{sp}.

With these derivation, we know it is right to use the conclusions of Sec. \ref{sp} and Sec. \ref{gw}. The quantum potential will have impact on the background wave function. It can give the relation between the scale factor and the Hubble parameter, which will determine the perturbations.
 
\section{Experimental constraints of vacuum inflation} \label{ravtm}

There are two parameters in the vacuum inflation model that should be determined: $C_1/C_2$ and $p$.
Only when all parameters are determined, can this model be falsified by experimental observations.
With the progress of measuring CMBR, this model can be fixed by observational results.

There are some parameters that constrain inflation in the current observations:
the tensor-to-scalar ratio $r\equiv\mathcal{P}_{g}/\mathcal{P}_{\mathcal{R}}^\prime<0.11(95\%\text{CL})$,
the spectral index of curvature perturbations $n_s\equiv1+d \mathcal{P}_{\mathcal{R}}^\prime/d\text{ln} k=0.968\pm0.006$
and its scale dependence $d n_s/d\text{ln} k=-0.003\pm0.007$ ~\cite{Ade2015lrj}.
From Eq.~(\ref{pr}) and Eq.~(\ref{pg}), we express these parameters up to the second-order of HFFs:
\begin{align} \label{cs}
\begin{split}
r&=8 \epsilon_1\left( 1+\epsilon_1+\left(C+1\right) \epsilon_1 \epsilon_2\right) \\
n_s-1&=-2 \epsilon_1- \epsilon_2-2 \epsilon_1^2-(2C+3) \epsilon_1 \epsilon_2 -C \epsilon_2 \epsilon_3\\
d\, n_s/d\,\text{ln}\, k&=-2 \epsilon_1 \epsilon_2- \epsilon_2 \epsilon_3,
\end{split}
\end{align}
where $C=-2+\text{ln}2+ \gamma$. Because the perturbations are mainly produced in the early stage of inflation,
we calculate the fluctuations when the scale factor is small. At that time, the influence of matter is negligible.
In this case, Eq.~(\ref{H}) gives
\begin{align} \label{ep}
&\epsilon _1=p+2
 &\epsilon _i=0 \big\lvert _{i>1} .
\end{align}
Then, Eq.~(\ref{cs}) can be represented by the ordering factor $p$ as $r=8(p+2)(p+3)$, $n_s-1=-2(p+2)(p+3)$
and $d\, n_s/d\,\text{ln}\, k=0$.

The measurement precision of $n_s$ is more dependable among the constraints of inflation, so we choose it to determine $p$, and get
$p+2=0.016\pm0.003$. If we choose
\begin{equation}
p+2=0.013,
\end{equation}
the limitation on $r$ can also be fixed. The restriction on $d\, n_s/d\,\text{ln}\, k$ is satisfied
regardless of the value of $p$. The $r$ and $n_s-1$  are constant in
the early stage of inflation, which is similar to slow-roll inflation.

To constrain the parameter $C_1/C_2$, we need to specify the starting point problem of scale factor. In this model, we can not give the starting point of $a$. Luckily enough, with Eq.~(\ref{h}), we can roughly determinate the ending point of the inflation by comparing the quantum potential with the contribution of matter. Once the ending point of $a$ is determined, we can use it to fix the starting point of $a$ with the general requirement of 60 $e$-folds of expansion~\cite{Lidsey:1995np}. Generally, the observational perturbations are mainly crossed the Hubble radius in the range from 50 to 60 $e$-foldings before the end of inflation~\cite{Lidsey:1995np}. 

With these general considerations, from Eq.~(\ref{pr}) and Eq.~(\ref{ep}), we can get
\begin{equation} \label{prp}
\mathcal{P}_{\mathcal{R}}^\prime=\frac{(1-(C+1) (p+2))^2}{8 (p+2) (p+3)}\frac{16}{\pi}H^2.
\end{equation}
The observations give that $\mathcal{P}_{\mathcal{R}}^\prime\approx2\times10^{-9}$ ~\cite{Ade2015lrj}. Before we constrain the parameter $C_1/C_2$ with this observation result, we still have an unsolved problem.
We can not numerically compute the ending point of the inflation without parameter $C_1/C_2$. Since $p+2$ is a very small number, $\mathcal{P}_{\mathcal{R}}^\prime$ is a slowly varying function of $a$. The estimation of the end point of inflation is pretty accurate. Using $a=1$ in Eq.~(\ref{H}) and Eq.~(\ref{prp}), we can get $C_1/C_2\approx1.55\times10^{-6}$. By numerical calculation, we can roughly get the end point of inflation around $a\approx10^4$. This means the observational perturbations are mainly crossed the Hubble radius in the range from $a=10^{-22}$ to $a=2\times10^{-18}$. With Eq.~(\ref{H}), Eq.~(\ref{prp}) and this range, we can rederive $C_1/C_2$ more accurately, which ranges from $C_1/C_2=8.0\times10^{-7}$ to  $C_1/C_2=9.2\times10^{-7}$. The range is really narrow and we set 
\begin{equation} \label{c1c2}
C_1/C_2=8.6\times10^{-7}
\end{equation}
for simplification. With this value of $C_1/C_2$, we can recompute the ending point of inflation. It changes within a magnitude of $10$, which proves that the estimation for $C_1/C_2$ is pretty good.
Thus, all the parameters in the vacuum inflation model have been determined.

\section{COMPARISON AND PREDICTION} \label{PC}
With the fixed parameters $C_1,C_2$ and $p$ , we will first calculate the matter generation. Then, we can test the model by comparing the calculated matter density with current experimental observations.

\begin{figure}[ht]
  \centering
    \includegraphics[width=0.45\textwidth]{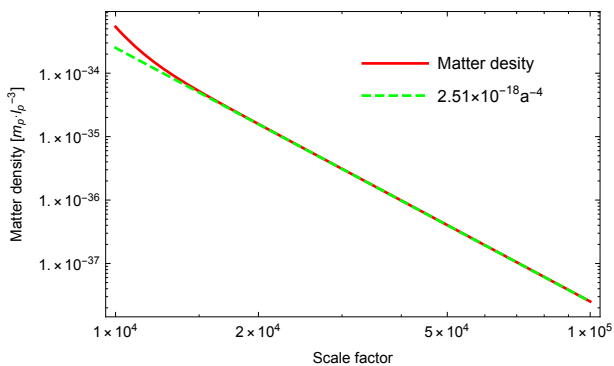}
	\caption{There are two curves in this figure: the energy density in our model and its fitting curve with the function proportional to $a^{-4}$. The figure uses the parameter fixed in Sec. \ref{ravtm}. We can see that the evolution of energy density coincide with the picture of radiation-dominated stage after $a\sim2\times10^{4}$. }\label{pmd}
\end{figure}

The evolution of the early universe is roughly determined by Eq.~(\ref{h}) and Eq.~(\ref{rho}). Before we do any calculation, we must clarify the evolution process of universe in this model. As all the parameter have been determined, we can divide
 the evolution process into several stages. At first, the cosmos is beginning to inflate due to quantum potential.
The scale factor grew very fast from about $10^{-22}$ to about 50, which is obvious in Fig.~(\ref{cmqh}). In this stage, matter is produced with little influence on the inflation.
With the growth of the scale factor,  the quantum potential decreases rapidly. It means that the universe  is about to exit the
inflation. This process lasts from about 50 to the magnitude of $10^4$, which can be seen from Fig.~(\ref{cmqh}), Fig.~(\ref{cm}) and 
Fig.~(\ref{pmd}). After the end of inflation, the radiation produced by inflation takes the role
of the quantum potential and drives the expansion of the cosmos slowly.
Combining Eqs.~(\ref{rho}) and (\ref{h}) and considering that $H$ is very small in this stage,
we have the relation
\begin{equation}
H\propto \frac{1}{a^2}
\end{equation}
The quantum potential is negligible at this stage, which is shown in Fig.~(\ref{pmd}).
After the radiation-dominated stage, the matter produced during inflation replaces the radiation and continues to drive the expansion of the cosmos.

Because the Hubble parameter grew rapidly during the initial stage of inflation and slowly after that stage,
we divide the calculations into two parts to obtain more accurate results. First, we plot the
function $H(a)$ to find when the quantum potential becomes negligible.
Fig.~(\ref{pmd}) shows that the quantum potential and matter production no longer
play a part when $a$ is very large.
The relation between $H$ and $a$ exactly fits the picture of the radiation-dominated stage.
In this case, we can ignore the effects of the quantum potential.

When the quantum potential dominates the inflation, we use
\begin{equation} \label{dt}
t_f-t_i=\int_{a_i}^{a_f} \frac{1}{\text{aH}} \, d\,a
\end{equation}
to evaluate the time $t_f$, we can use $a=2\times10^4$ from Fig.~(\ref{pmd}).
We use the time when $a=1$ as $t_i$. The reason why we use $t_i$ to calculate $t_f$ is that we have to calculate
this period with sufficiently long step length. The scale factor $a=1$ does not mark the start of the universe as we clarified before. 
If we choose $a=10^{-22}$ at the start of time, $t_i$ will be a very short time. The universe is still inflating at $t_i$ according to Fig.~(\ref{cmqh}).
The Hubble parameter can be described by Eq.~(\ref{H}) at this time. Therefore, we have
\begin{equation}
t_i=\int_{10^{-22}}^1 \frac{2}{l^2}\frac{C_2 }{C_1} a^{-p-3}\, da\approx1.0\times10^7\, t_P,
\end{equation}
where $t_P$ is the Planck time. Using Eqs.~(\ref{h}), (\ref{qp}) and (\ref{rho}), we can obtain the evolution of the Hubble parameter
along with the scale factor numerically. With the numeric solution of the Hubble parameter we can integrate Eq.~(\ref{dt}) numerically. We calculate $t_f-t_i\approx3.40\times10^{16}\, t_P$,
which is much bigger than $t_i$. The matter density can be  calculated as $\rho_f\approx1.58\times10^{-35}\,m_P\cdot l_P^{-3} $. We now address the classic case. With the Friedmann equation $H^2=8\pi\rho /3$ and the mass-conservation equation $\dot{\rho }=-3  (w+1)H \rho $ we can obtain the evolution of matter density
\begin{equation} \label{hc}
\rho\propto t^{-2},
\end{equation}
where $w$ represents the ratio of the pressure to the density of matter, which we treat as constant because it changes very slowly. Before we use Eq.~(\ref{hc}), we must be careful about the meaning of $t$. $t=0$ does not mark the time when $a=10^{-22}$. We must calculate asymptomatic axis where $\rho \to \infty$ from Eq.~(\ref{hc}) and Fig.~(\ref{pmd}).
We can get $t_a\approx-0.95\times10^{16}\, t_P$ by comparing $\rho$ and $t$ of points after $a=2\times10^4$. The right relation can be get by replacing $t$ with $t-t_a$ in Eq.~(\ref{hc}).
The age of the universe is estimated as $t_{now}\approx4.3\times10^{17}\,s$ ~\cite{Ade:2013sjv}. Using revised Eq.~(\ref{hc}), we estimate
\begin{equation}
\rho_{now}=\rho_f(\frac{t_f-t_a}{t_{now}})^2\approx2.5\times10^{-27}\, \text{kg}\cdot \text{m}^{-3},
\end{equation}
 which is roughly in agreement with experimental observation, which gives $\rho\approx1.8\times10^{-27}\, \text{kg}\cdot \text{m}^{-3}$ ~\cite{Ade:2015xua}.

The model can also provide some new predictions for further test. The HFFs of this model are different from other models. The high-order terms of the HFFs are zero as shown in the Eq.~(\ref{ep}).
Independent of the value of $p$, we have two strict relations 
\begin{align}
	\begin{split}
		r/(1-n_s)=4,\\
		n_s/d\,\text{ln}\, k=0
	\end{split}
\end{align}
These can be used to falsify this model in future, when more accurate observations are available.


\section{DISCUSSION AND CONCLUSION}

The vacuum inflation model derived by the quantum potential is interesting because
it can avoid the problem of \emph{creation}~\cite{edp73}.
In contrast to other inflation models,
the early universe is a vacuum bubble that grows rapidly due to its quantum potential.
Matter can be created during the inflation stage, and inflation ends when the
early universe becomes large enough.
In this paper, we have considered the scalar perturbations and gravitational waves
for the vacuum inflation derived by the quantum potential with HFFs directly.
We find that matter contributes little to inflation, but becomes significant after the inflation.
All the parameters in the vacuum inflation model are determined by comparison with observations. The matter density of the current universe is calculated with the
vacuum inflation model, and the result roughly fits the observations, which indicates that the vacuum
inflation model may be the correct answer.

\section{ACKNOWLEDGMENTS}
We thank the referees for their helpful comments and
suggestions that significantly polish this work. This work is supported by the National Natural Science
Foundation of China (Grant Nos. 61471356).

\end{document}